\documentclass[aps,prd,twocolumn,superscriptaddress,nofootinbib]{revtex4}

\usepackage{graphicx}
\usepackage{amsmath,amssymb,latexsym}
\usepackage{bm}
\usepackage{epsfig}
\newcommand{\postscript}[2]{\setlength{\epsfxsize}{#2\hsize}
   \centerline{\epsfbox{#1}}}

\newcommand{\barr}{\begin{array}}
\newcommand{\earr}{\end{array}}
\newcommand{\bea}[1]{\begin{eqnarray} \label{(#1)}}
\newcommand{\eea}{\end{eqnarray}}
\newcommand{\beas}{\begin{eqnarray*}}
\newcommand{\eeas}{\end{eqnarray*}}
\newcommand{\ca}{{\cal A}}
\newcommand{\ma}{M(\ca)}
\newcommand{\eav}{\langle E_\nu\rangle}
\newcommand{\ncritnr}{n_{\nu,{\rm c}}^{\rm NR}}
\newcommand{\ncritrel}{n_{\nu,{\rm c}}^{\rm REL}}
\newcommand{\ed}{\end{document}}

\def\lsim{\mathrel{\raise.3ex\hbox{$<$\kern-.75em\lower1ex\hbox{$\sim$}}}}
\def\gsim{\mathrel{\raise.3ex\hbox{$>$\kern-.75em\lower1ex\hbox{$\sim$}}}}

\begin{document}

\title{Phase transition in the fine structure constant}

\author{L.~Anchordoqui}
\affiliation{Department of Physics, University of Wisconsin-Milwaukee,
Milwaukee, WI 53201
}

\author{V.~Barger}
\affiliation{Department of Physics,
University of Wisconsin,
Madison, WI 53706
}

\author{H.~Goldberg}
\affiliation{Department of Physics,
Northeastern University, Boston, MA 02115
}

\author{D.~Marfatia}
\affiliation{Department of Physics and Astronomy,
University of Kansas,
Lawrence, KS 66045
}

\begin{abstract}

  \noindent
 
Within the context of mass-varying neutrinos, we construct a
  cosmological model that has a phase transition in the 
electromagnetic fine
structure constant $\alpha$ at a redshift of 0.5.  
The model accommodates hints of a time variable  
$\alpha$ in quasar spectra and the nonobservance of such an 
effect at very low redshifts.
It is consistent with limits from the recombination and primordial
  nucleosynthesis eras and is free of instabilities.


\end{abstract}


\maketitle

\section{Introduction}

Measurements of the cosmic microwave background, large scale
structure, the evolution of the Hubble parameter from
luminosity-redshift relation of type Ia 
supernovae along with the abundances of
light elements in the universe  strongly
indicate the existence of a dark energy of unknown origin that acts
against
the pull of gravity~\cite{Copeland:2006wr}.  
The combined data favor an effective de-Sitter constant
that nearly saturates the upper bound given by the present-day
 value (which we denote by a subscript 0 to indicate redshift $z=0$) 
of the Hubble 
parameter $H_0 \approx 10^{-33}$~eV. This yields a
dark energy density: $\rho_{\rm DE} \sim 3 M_{\rm Pl}^2 H_0^2 \sim
(2.4 \times 10^{-3}~{\rm eV})^4,$ where $M_{\rm Pl} \simeq 2.4 \times
10^{18}~{\rm GeV}$ is the Planck mass.

The coincidence of the neutrino mass scale with the dark energy mass
scale
is suggestive that there may be a link between these quantities.
Measurements of atmospheric neutrinos have provided evidence (at $>
15\sigma$) for $\nu_\mu$ disappearing (likely converting to
$\nu_\tau$) when propagating over distances of order hundreds (or
more) kilometers. The corresponding oscillation
phase is consistent with being maximal and the oscillations require a
neutrino mass-squared difference
of $\delta m^2_{\rm atm} \sim 2.5 \times 10^{-3}$~eV$^2$~\cite{Barger:2003qi}.
The  $\nu_\mu$ disappearance oscillations have been confirmed by the
KEK-to-Kamioka (K2K)  and MINOS experiments over baselines of
250~km  and 730~km, respectively.
To implement a connection of dark energy and neutrino mass~\cite{pq} in a
concrete manner,  Fardon, Nelson, and
Weiner (FNW)~\cite{Fardon:2003eh} introduced a Yukawa coupling between
a sterile neutrino and a cosmic scalar field (dubbed the acceleron),
such that the neutrino masses $m_{\nu_i}$ are generated by the vacuum
expectation value ${\cal A}$ of this field, {\it i.e.}, $m_{\nu_i} ({\cal
  A})$.  For simplicity hereafter we only consider a single
nonvanishing neutrino mass, $m_\nu$. The active
neutrino mass is determined through a seesaw mechanism by
integrating out the heavy sterile
neutrino with mass $M(\ca)$, now correlated with the acceleron.  This
gives an effective potential
\begin{eqnarray}
V^{\rm NR}_{\rm eff} &  = &  m_\nu (\ca)\ n_\nu + V[\ma] \nonumber \\
            & = &  {m_D^2\over \ma}\ n_\nu\ + V[\ma]
\label{VNR}
\end{eqnarray}
for regions in which nonrelativistic neutrinos
dominate~\cite{Fardon:2003eh} and
\begin{eqnarray}
V^{\rm REL}_{\rm eff} &  = &  m_\nu (\ca)^2\ {n_\nu\over \eav} + V[\ma]
\nonumber \\
            & = & {m_D^4 \over \eav \ma^2}\ n_\nu + V[\ma]
\label{VREL}
\end{eqnarray}
for regions where relativistic neutrinos dominate.  Here $V$ is the
fundamental acceleron potential, $m_D$ is a
Dirac neutrino mass and $\eav$ is the average (relativistic) neutrino
energy.  At the minimum of the potential, the neutrino mass is
determined in terms of its density $n_\nu$. This creates neutrino mass
dependence on the environment with possible relevance to 
solar~\cite{Barger:2005mn} and short-baseline neutrino
oscillations~\cite{Barger:2005mh},
as well as the cosmic neutrino background~\cite{Ringwald:2006ks}. 
Interestingly, since the
``constants'' of the Standard Model depend nontrivially on the scalar
neutrino density, $m_{\nu} ({\cal A})$ could induce variations in the
fine structure constant of quantum electrodynamics, $\alpha$, which
is the focus of this Letter.

The fine structure constant has been measured in the spectra of
distant quasars (QSO) for a number of absorption systems.
Early high redshift measurements of $\alpha$ with the Keck telescope 
found no discrepancy in
comparison with laboratory measurements of $\alpha$ to an accuracy of a
few parts in $10^{-4}$~\cite{Cowie:1995sz} 
and other observers also subsequently put upper
limits on any discrepancy below the $10^{-5}$ level.  However, a discrepancy
of $\Delta\alpha/\alpha = -0.57\pm 0.10\times 10^{-5}$ was reported
in Ref.~\cite{Webb:1998cq}.  Further observations with 
the Very Large Telescope found no
discrepancy at this level~\cite{Srianand:2004mq}, 
but the parameter estimation methods are currently under
 debate~\cite{Murphy:2006vs}.  Even if a discrepancy exists,
it is not excluded that an effect may be imitated by a
large change of isotope abundances over the last 10 billion 
years~\cite{Ashenfelter:2003fn}. Thus further observations are mandated to
definitively decide whether or not $\alpha$ is truly constant.

It has been proposed that
a variation of $\alpha$ could result from the temporal evolution
of a quintessence field~\cite{Anchordoqui:2003ij}.
However, the model predicts a rather small variation of $\alpha$ from
high redshifts to the present unless the quintessence field has
unexpectedly undergone a rapid slowing in the recent past.

Here we pursue the implications of mass-varying neutrinos (MaVaNs) on
the variation of $\alpha$ over cosmic time scales. For a class of
dependences of $\ma$ and $V$ a
transition in the neutrino phase may occur as the neutrino density
varies~\cite{Afshordi:2005ym}. We show that this phase transition allows the
existence of
two distinct stable phases for $\alpha$.{\footnote{For a completely
different mechanism that leads to an abrupt change in $\alpha$, see
Ref.~\cite{maxim}.}

Our paper is organized as follows.  In Sec.~\ref{2} we analyze the
requirements that a MaVaN
phase transition occurs and discuss the conditions under which the
model can
circumvent hydrodynamic instabilities that may be manifest in the
nonrelativistic regime~\cite{Afshordi:2005ym,Takahashi:2005kw}. Then,
in Sec.~\ref{3}, we study the corresponding implications for the
time variation of the fine structure constant. We
summarize in Sec.~\ref{4}.

\section{Phase Transition in the Mass Variation of Neutrinos}
\label{2}

The stationary points for the potentials in Eqs.~(\ref{VNR}) and
(\ref{VREL}) are given by
\begin{equation}
\frac {dV^{\rm NR}_{\rm eff}}{d\ca} =
\left( -\frac{m_D^2\ n_\nu}{M^2} + V^{\prime}(M)\right)
\frac{dM}{d\ca}\ =\ 0
\label{statnr}
\end{equation}
and
\begin{equation}
\frac {dV^{\rm REL}_{\rm eff}}{d\ca} =
\left( -\frac{2 m_D^2\ n_\nu}{\eav\ M^3} + V^{\prime}(M)\right)
\frac{dM}{d\ca}\ =\ 0 \, ,
\label{statrel}
\end{equation}
where $V^{\prime}(M)\equiv \partial V(M)/\partial M$ in the two
cases.  We choose the phase of the singlet neutrino field so that $M$
is real and nonnegative.

To examine the possibilities for a multiphase structure, we note that
 in string-based discussions masses and
couplings are determined by the minima of stabilized
moduli~\cite{Damour:1994zq}. For our purposes, we take this to mean that
$M(\ca)\simeq M_o [ 1 + (\ca - \ca_o)^2/f^2]$ in the vicinity of the
stabilization point $\ca_o,$ where $f$ is a positive constant. Without loss of
generality, we set $\ca_o=0$. With $M_o\ne 0,$ we make
the assumption: (I) $M(\ca)$ has a unique stationary point at its
absolute minimum $M_o$.  (Although not essential for the discussion,
we adopt the simplifying assumption that $M$ is an even function
of $\ca$.) As a consequence, both $V^{\rm NR}_{\rm eff}$ and $V^{\rm
  REL}_{\rm eff}$ have stationary points at $\ca=0$ where $dM/d\ca=0$. From
Eqs.~(\ref{statnr}) and (\ref{statrel}) additional stationary
points will exist if the following permits solutions:
\begin{equation}
\left(\frac{M}{M_o}\right)^j\;
\frac{V^{\prime}(M)}{V^{\prime}(M_o)}
= \frac{n_\nu}{{n_{\nu,{\rm c}}^{i}}}\,,
\label{simpnr}
\end{equation}
where $j=2,3$ if $i={\rm NR, REL}$
for the nonrelativistic and relativistic cases, respectively. Here
\begin{equation}
\ncritnr \equiv {M_o^2\over m_D^2}\ V^{\prime}(M_o)\,, \ \ \ \ 
\ncritrel \equiv {\eav\ M_o^3\over 2\,m_D^4}\ V^{\prime}(M_o)\,.
\label{ncrit1}
\end{equation}
To proceed, rather than examining the system in full generality, we
prefer to illustrate the possibilities by imposing a further
condition, namely (II) $M^2 V^{\prime}(M)$ is an increasing
function of $M$. With the help of conditions I and II,
Eq.~(\ref{simpnr}) will have solutions in their
separate domains if and only if $n_\nu > n_{\nu, {\rm c}}$. If this
condition is fulfilled, the additional stationary points are the two
mirror values $\pm \ca_{\rm min}$ corresponding to the value $M$
(assuming there is only one) for which there is a solution.  These
arguments then imply that the effective potential has the form in
Fig.~\ref{potential}, where the lower (upper) curve is valid for
$n_{\nu}< n_{\nu_{\rm c}}$ ($n_{\nu} > n_{\nu_{\rm c}}$).

\begin{figure}
\begin{center}
\postscript{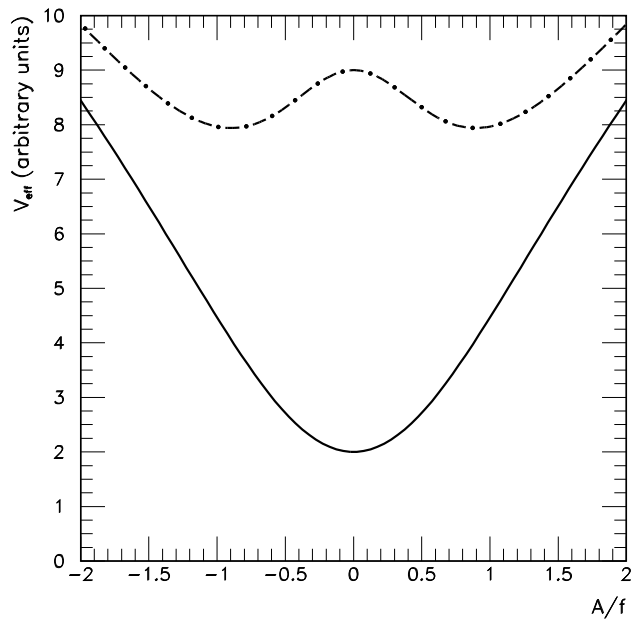}{0.98}
\caption{Qualitative behavior of the effective potential. The
  dot-dashed line indicates the supercritical regime where 
 \mbox{$n_{\nu} > n_{\nu_{\rm c}}$}, 
 whereas the solid line indicates the
  subcritical regime where $n_{\nu}< n_{\nu_{\rm c}}$. }
\label{potential}
\end{center}
\end{figure}

To illustrate these considerations, and to show how to circumvent the
instability issues raised in Ref.~\cite{Afshordi:2005ym} in the
nonrelativistic regime, we choose a slight variant of the original FNW
form~\cite{Fardon:2003eh} for the acceleron potential,
\begin{equation}
V[\ma] = \Lambda^4\ \ln (|\ma/M_o|)\,\, ,
\label{wn}
\end{equation}
normalized so that $V=0$ at $\ca=0.$
Equation~(\ref{simpnr}) becomes
\begin{equation}
\left(\frac{M}{M_o}\right)^k = \frac{n_\nu}{n_{\nu,{\rm c}}^i}\,,
\label{simprellog}
\end{equation}
where $k=1,2$ for $i={\rm NR, REL}$, respectively. 
The critical neutrino densities may be calculated using
Eq.~(\ref{ncrit1}). In terms of
the mass of the
heaviest neutrino in a dilute environment
($m_{\nu,0} \equiv m_D^2/M_o
\agt \sqrt{\delta m^2_{\rm atm}}$), we find
\begin{equation}
\label{ncritnrel}
\ncritnr =
\frac{\Lambda^4}{m_{\nu,0}}\,, \ \ \ \ \ \ \
  \ncritrel = \frac{\eav\ \Lambda^4}{2\,m_{\nu,0}^2}\simeq 
\frac{T_\nu\ \Lambda^4}{m_{\nu,0}^2 }\,\,.
\end{equation}
Now suppose that for nonrelativistic $\nu + \overline \nu$, the
neutrino density is subcritical. Then the only stationary point is at
$M=M_o\ (\ca = 0)$ yielding a neutrino mass $m_\nu = m_{\nu,0}$, which
is {\em independent of the neutrino density}, so that there are {\em
  no stability problems.}  Instability due to increasing
densities can occur at a redshift at which both (a) $n_\nu >
n_{\nu, {\rm c}}$ and (b) the neutrinos are nonrelativistic, with
temperature $T_\nu \lsim 1.2\, \Lambda $~\cite{Afshordi:2005ym}. However,
recent work~\cite{Bjaelde:2007ki} has shown that the instability
 can be avoided for sufficiently weak coupling of the neutrinos
to the acceleron during the relevant cosmological era. 
We explicitly show below that there is indeed a window of late-time
instability for the FNW model~\cite{Fardon:2003eh}, but that it can be
avoided if the acceleron couplings are gravitational or stringy in
origin.

In the nonrelativistic regime, and for $n_\nu > n_{\nu,{\rm c}}^{\rm NR}$,
from Eqs.~(\ref{simprellog})-(\ref{ncritnrel}) we have
$m_\nu = \Lambda^4/n_\nu$ with $n_\nu=3 \zeta(3)/(2 \pi^2) T_\nu^3$.
The neutrinos will be nonrelativistic if
\begin{equation}
\frac{\sqrt{\langle p_{\nu}^2\rangle}}{m_\nu} =
\sqrt{\frac{15\zeta(5)}{\zeta(3)}}\frac{T_\nu n_\nu}{\Lambda^4} < 1\,,
\end{equation}
where $\langle p_{\nu}^2\rangle$ is the mean square neutrino momentum
in the Fermi-Dirac distribution. 
This yields
$T_\nu\lsim 1.1\Lambda$,
which effectively coincides with the criterion for instabilities
in Ref.~\cite{Afshordi:2005ym}. 
With the condition $n_\nu > \ncritnr$, the window of instability is 
\begin{equation}
1.8  \left(\frac{\Lambda}{m_{\nu,0}}\right)^{1/3} \lsim \frac{T_\nu}{\Lambda} 
\lsim 1.1 \,\,.
\label{five}
\end{equation}
Thus, instabilities may appear if $\Lambda/m_{\nu,0} \lsim 0.23$.

Since $T_\nu = T_{\nu,0} (1+z)$  (with $T_{\nu,0}
\simeq 1.7 \times 10^{-4}$~eV), Eq.~(\ref{five}) can be expressed in 
terms of redshift as
\begin{equation}
 2.9 \left(\frac{\Lambda_{-3}^4}{m_{\nu,0}/0.05~{\rm eV}}\right)^{1/3} \lsim 
1+z \lsim 6.5 \  \Lambda_{-3} \,\, ,
\label{four}
\end{equation}
where $\Lambda_{-3} \equiv \Lambda/(10^{-3}\ {\rm eV})$.

Instabilities may be avoided if the coupling between the acceleron and 
neutrinos $\beta$ satisfies the inequality~\cite{Bjaelde:2007ki}
\begin{equation}
\beta \equiv \left| \frac{d\ln m_\nu}{d {\cal A}}\right| =
\left|\frac{d \ln M}{d{\cal A}} \right| < \sqrt{\frac{\Omega_{\rm CDM} - \Omega_\nu
}{2\, \Omega_\nu}} \, \frac{1}{M_{\rm Pl}} \,\,,
\label{beta}
\end{equation}
where $\Omega_{\rm CDM}$ ($\Omega_\nu$) is the cold dark matter
(neutrino) dimensionless density. For values of interest, $\beta < 10/M_{\rm
  Pl}$~\cite{Bjaelde:2007ki}. So far our analysis has been
independent of a particular choice of $\ma;$ however, a test of this
criterion, and in what follows, the results are somewhat dependent on
this choice.  We consider two simple forms (which also satisfy
assumptions I and II above): 
\begin{equation}
M = M_o \, e^{{\cal A}^2/f^2}
\label{gauss}
\end{equation}
and 
\begin{equation}
M = M_o\, \cosh{{\cal A}/f}\ \ ,
\label{cosh}
\end{equation}
reminiscent of M-theory potentials for moduli~\cite{Cvetic:1999xp}.
We analyze the first case in detail, and only provide results for the second.
From Eq.~(\ref{beta}),
\begin{equation}
\beta = \frac{2|\ca|}{f^2} < \frac{10}{M_{\rm Pl}} \,\,.
\label{opo}
\end{equation}

A bound on $|\ca|/f$ may be obtained by imposing the
nonrelativistic criterion $T_\nu/\Lambda < 1.1$. From
Eqs.~(\ref{simprellog}) and (\ref{ncritnrel}),
\begin{equation}
e^{\ca^2/f^2} = 
\frac{3 \zeta(3)}{2 \pi^2} \, {m_{\nu,0}\over \Lambda}  \,
\left(\frac{T_\nu}{\Lambda}\right)^3 \,\,,
\end{equation}
which gives
\begin{equation}
\frac{|\ca|}{f}< \sqrt{ \ln\frac{12 \,\,(m_{\nu,0}/0.05~{\rm eV})} {\Lambda_{-3}} }
\, .
\end{equation}
In our discussion of the $\alpha$ variation we will find that
$\Lambda_{-3} \simeq 0.6 (m_{\nu,0}/0.05~{\rm eV})^{1/4},$ so that for
$m_{\nu,0}$ not much in excess of 0.05~eV, $|\ca|/f < 1.7$. From
Eq.~(\ref{opo}) we see that $f/M_{\rm Pl} > 0.34$ serves as a sufficient
condition to avoid instabilities.

For the alternate $\ma$ of Eq.~(\ref{cosh}), we find
$\beta=|\tanh(\ca/f)|/f \le 1/f$ for all $\ca$, so that
$f/M_{\rm Pl} > 0.1$ provides a sufficient condition for weak coupling
and stability.


\section{Discontinuity in the Fine Structure Constant}
\label{3}

Allowing for couplings between the acceleron and standard model 
fields,{\footnote{We do not prescribe a mechanism (which may be desirable 
for technical naturalness) via which loop corrections 
to the acceleron potential are suppressed.}} 
the free Lagrangian for the electromagnetic field tensor
$F_{\mu\nu}$ can be written as
\begin{equation}
 \widetilde{\cal L}_{\rm em}
= -\frac{1}{4}\ Z_F({\cal A}/M_{\rm Pl})\ F_{\mu\nu}F^{\mu\nu}\,,
\end{equation}
which on expansion about the present value ${\cal A}_0$ of ${\cal A}$, becomes
\begin{equation}
\widetilde{\cal L}_{\rm em} = -\frac{1}{4}\
(1+ \kappa \ \Delta{\cal A}/M_{\rm Pl} \ + \ldots) \,
F_{\mu\nu}F^{\mu\nu}\ \ ,
\label{expansion}
\end{equation}
with $\Delta{\cal A} = {\cal A}- {\cal A}_o$
and $\kappa\equiv \left. \partial_\ca Z_F \right|_{{\cal A}_o}$. The field
renormalization $A_{\mu}\rightarrow A_{\mu}/Z_F^{1/2}$ to obtain
a canonical kinetic energy, generates  an effective charge
$e/Z_F^{1/2}$. Following Ref.~\cite{Anchordoqui:2003ij}, we expand to linear order
about the present value
$e_0$, to obtain
\begin{equation}
\left|\frac{\Delta \alpha}{\alpha}\right| = \kappa
\frac{\Delta{\cal A}}{M_{\rm Pl}} = \kappa\ \frac{{\cal A}}{f}\,\cdot \frac{f}{M_{\rm Pl}}\ \ ,
\label{alpha}
\end{equation}
where $\alpha \equiv e^2/(4\pi)$. Equation~(\ref{alpha}) reflects our 
assumption that the variation
in $\alpha$ is uniquely derived from the evolution of the acceleron.

In order to accommodate the meteorite data~\cite{smoliar}, which do not
show evidence for a time-dependent $\alpha$, we require that
${\cal A}$ not vary from
 ground state equilibrium $(\ca_o = 0)$ for $z \lsim 0.5$. 
Consequently, the model predicts no
variation of $\alpha$ during this era, in agreement with existing
limits~\cite{Olive:2002tz}. 

From Eq.~(\ref{four}), for a transition at $z = 0.5$,
\begin{equation}
\Lambda_{-3} \simeq 0.61 (m_{\nu,0}/0.05~{\rm eV})^{1/4}\,,
\label{Lambda}
\end{equation}
or equivalently
\begin{equation}
  \frac{\rho_{\cal A}}{\rho_{\rm DE}} \sim
4 \times 10^{-3} \,\, \frac{m_{\nu,0}}{0.05~{\rm eV}}\,.
\end{equation}
This precludes $\Lambda$ saturating the present dark energy. This
ratio is roughly similar to the neutrino contribution 
to the dark matter density. 
Large scale surveys and WMAP together constrain the neutrino
energy density to be $\Omega_\nu
\alt 0.02,$ whereas terrestrial measurements of the neutrino mass
indicate $\Omega_\nu > 7 \times 10^{-4}$.

From Eqs.~(\ref{four}) and~(\ref{Lambda}), the density is supercritical and the 
neutrinos are nonrelativistic for $0.5 <z \lsim 4$. 
So there will be a tiny variation of $\alpha$, in agreement with
observations of absorption lines in the spectra of distant
QSO~\cite{Webb:1998cq}. The $z$-dependence of this variation may  be obtained
in a straightforward manner.  From Eq.~(\ref{simprellog}), at the minimum,
\begin{equation}
\frac{M(\ca)}{M_o} = \left( \frac{1+z}{1+z_c}\right)^3\ \ ,
\label{Mz}
\end{equation}
where $z_c$ ($\simeq 0.5$ in our case) is the redshift for which $n_\nu
= \ncritnr.$ For the gaussian potential, we have
\begin{equation}
\frac{|\ca|}{f} = \sqrt{3\ \ln\left(\frac{1+z}{1+z_c}\right)}\ \ ,
\label{Azgauss}\
\end{equation}
and for the cosh potential,
\begin{equation}
\frac{|\ca|}{f} = \cosh^{-1}\left[\left(\frac{1+z}{1+z_c}\right)^3\right]\ \ .
\label{Azcosh}
\end{equation}
At $z=2$ (the intermediate point of the data), $|\ca|/f\simeq 1.4 \, (2.8)$
for the gaussian and cosh cases, respectively. Taking $f/M_{\rm Pl}\ge 0.34\,
(0.1)$ for the two cases from the previous section, we find from Eq.~(\ref{alpha}),
\begin{eqnarray}
\left| \frac{\Delta \alpha}{\alpha}\right| & \gsim &  0.5\  \kappa \qquad \qquad {\rm gaussian}\\
\left| \frac{\Delta \alpha}{\alpha}\right| & \gsim &  0.3\  \kappa \qquad \qquad {\rm cosh}
\label{daa}
\end{eqnarray}
Thus,  accommodating the possible variation of $\alpha$ at a level
of 5 parts per million requires $\kappa \sim 10^{-5}.$

The quantity $\kappa$ may be bounded by available limits on
$\Delta \alpha/\alpha$ during the eras of recombination $(z\simeq
1100)$ and big bang nucleosynthesis (BBN) $(z\sim 10^{10})$. Since  
neutrinos are relativistic at these redshifts, there are no stability
problems. It is straightforward to show that $n_\nu > \ncritrel$
as soon as the neutrinos become relativistic. Then, the system is
in the dot-dashed phase of Fig.~\ref{potential}, and at the (mirror)
minima the field ${\cal A}$ is given by relativistic case of
Eq.~(\ref{simprellog}), 
\begin{eqnarray}
{M \over M_o} &= &
\sqrt{\frac{3 \zeta(3)}{2 \pi^2} \, 
\frac{T_\nu^2\ m_{\nu,0}^2}{\Lambda^4}}\,,
\label{afgauss}
\end{eqnarray}
which with Eq.~(\ref{Lambda}) yields
\begin{eqnarray}
|\ca|/f &\simeq & \sqrt{\ln\left(10\,z\right)} \qquad \ \ \ \ {\rm gaussian}\\
|\ca|/f &\simeq & \cosh^{-1} \left(10\,z\right) \qquad {\rm cosh}
\end{eqnarray}
Inserting these in Eq.~(\ref{alpha}), we find for the recombination and 
BBN eras,
\begin{equation}
\left| \frac{\Delta \alpha}{\alpha} \right| \simeq \frac{b_{\rm rec}\,\kappa\, f }{
M_{\rm Pl}} \,, \ \ \ \ \ \ \ \ \left| \frac{\Delta \alpha}{\alpha}\right| \simeq \frac{b_{\rm BBN}\,\kappa\, f }{M_{\rm Pl}} \,,
\label{recomb}
\end{equation}
where $b_{\rm rec}=3$, 10 and $b_{\rm BBN}=5$, 26 
for the gaussian and cosh potentials, respectively.
 (The slow variation with $m_{\nu,0}$ has been ignored in both cases.)
Existing limits on the
variation of the fine structure constant, $|\Delta \alpha/\alpha| \lsim 0.02$ 
(at the 95\%~C.~L.) for both recombination~\cite{Hannestad:1998xp} and
BBN~\cite{Kolb:1985sj}, in conjunction with the lower bounds on 
$f/M_{\rm Pl}$ can now be translated into
bounds on the coupling constant:
\begin{eqnarray}
\kappa &<& 0.02 \qquad {\rm recombination} \\
\kappa &<& 0.01 \qquad {\rm BBN}\ \,, 
\end{eqnarray}
for the gaussian potential and 
\begin{eqnarray}
\kappa & < & 0.02 \qquad \ \, {\rm recombination}\\
\kappa & < & 0.008 \qquad {\rm BBN} \,\,,
\end{eqnarray}
for the cosh potential.
The bounds on $\kappa$ are a few orders of magnitude larger than the
required value to accomodate existing data in the subrelativistic
regime.

\section{Summary}
\label{4}

In the MaVaN framework, we have constructed a cosmological model
that can accommodate limits on the variation of the fine
structure constant on short time scales, as well as the potential
observation of a variation in $\alpha$ from distant QSOs. The model
has a phase transition in the neutrino mass at a redshift
$z = 0.5$ that gives a phase transition in $\alpha$. The
existence of this phase transition precludes that the vacuum
energy associated with the acceleron field saturates the present dark
energy. 
To circumvent hydrodynamic instabilities we assumed a sufficiently weak
  coupling (perhaps stringy in origin) of the neutrinos to the
  acceleron during the cosmological evolution.  The model is
consistent with limits on
  $\Delta\alpha/\alpha$ from recombination and primordial
  nucleosynthesis.

\acknowledgments{We thank Lily Schrempp for valuable
  discussions.  LA is supported by the University of Wisconsin
  Milwaukee. VB is supported by the U.S. Department of Energy
  (DoE) under Grant No.  DE-FG02-95ER40896, and by the Wisconsin
  Alumni Research Foundation. HG is supported by the U.S. National
  Science Foundation (NSF) Grant No PHY-0244507. DM is supported by
  the DoE under Grant No. DE-FG02-04ER41308, and by the NSF under 
  CAREER Award No. PHY-0544278.}

\ed